\documentclass[12pt]{article}

\usepackage{amsmath}
\usepackage{stmaryrd}
\usepackage{subfigure}
\usepackage{empheq}
\usepackage{appendix,epic,eepic,amsmath,amsthm,amssymb,epsfig,cite,indentfirst,oldgerm}
\renewcommand{\theequation}{\arabic{equation}}
\newcommand{\EQ}{\begin{equation}}
\newcommand{\EN}{\end{equation}}

\newcommand{\IM}{\mathbf{\imath}}

\newcommand{\bear}{\begin{eqnarray}}
\newcommand{\ear}{\end{eqnarray}}
\newcommand{\bt} { \begin{tabular} }
\newcommand{\et}{ \end{tabular} }
\newcommand{\bc} { \begin{center} }
\newcommand{\ec}{ \end{center} }

\newcommand{\btb} { \begin{table} }
\newcommand{\etb}{ \end{table} }

\begin{document}

\topmargin 0pt
\oddsidemargin 5mm
\newcommand{\NP}[1]{Nucl.\ Phys.\ {\bf #1}}
\newcommand{\PL}[1]{Phys.\ Lett.\ {\bf #1}}
\newcommand{\NC}[1]{Nuovo Cimento {\bf #1}}
\newcommand{\CMP}[1]{Comm.\ Math.\ Phys.\ {\bf #1}}
\newcommand{\PR}[1]{Phys.\ Rev.\ {\bf #1}}
\newcommand{\PRL}[1]{Phys.\ Rev.\ Lett.\ {\bf #1}}
\newcommand{\MPL}[1]{Mod.\ Phys.\ Lett.\ {\bf #1}}
\newcommand{\JETP}[1]{Sov.\ Phys.\ JETP {\bf #1}}
\newcommand{\TMP}[1]{Teor.\ Mat.\ Fiz.\ {\bf #1}}

\renewcommand{\thefootnote}{\fnsymbol{footnote}}

\newpage
\setcounter{page}{0}
\begin{titlepage}
\begin{flushright}

\end{flushright}
\vspace{0.5cm}
\begin{center}
{\large Algebro-Geometric approach for a centrally extended $U_q[sl(2/2)]$ $\mathrm{R}$-matrix} \\
\vspace{1cm}
{\large M.J. Martins } \\
\vspace{0.15cm}
{\em Universidade Federal de S\~ao Carlos\\
Departamento de F\'{\i}sica \\
C.P. 676, 13565-905, S\~ao Carlos (SP), Brazil\\}
\vspace{0.35cm}
\end{center}
\vspace{0.5cm}

\begin{abstract}
In this paper we investigate the algebraic geometric nature of a solution of the 
Yang-Baxter equation based on the quantum 
deformation of the centrally extended $sl(2|2)$ superalgebra
proposed by Beisert and Koroteev \cite{BEKO}. We derive an 
alternative representation for the 
$\mathrm{R}$-matrix in which the matrix 
elements are given in terms
of rational functions depending on weights sited on
a degree six surface. For generic gauge the weights geometry are
governed by a genus one ruled surface while 
for a symmetric gauge choice the weights lie instead on a genus five curve.
We have written down the polynomial identities satisfied by
the $\mathrm{R}$-matrix entries needed 
to uncover the corresponding  
geometric properties. For arbitrary gauge the  $\mathrm{R}$-matrix geometry
is argued to be birational to the direct product
$\mathbb{CP}^1 \times \mathbb{CP}^1 \times \mathrm{A}$ where $\mathrm{A}$
is an Abelian surface. For the symmetric gauge we present evidences
that the geometric content is that of a surface of general type 
lying on the so-called Severi line with
irregularity two and geometric genus nine. We discuss potential geometric
degenerations when the two free couplings are restricted 
to certain one-dimensional subspaces.

\end{abstract}

\vspace{.15cm} \centerline{}
\vspace{.1cm} \centerline{Keywords: Yang-Baxter, $\mathrm{R}$-matrix, Algebraic Geometry}
\vspace{.15cm} \centerline{October 2016}

\end{titlepage}


\pagestyle{empty}

\newpage

\pagestyle{plain}
\pagenumbering{arabic}

\renewcommand{\thefootnote}{\arabic{footnote}}
\newtheorem{proposition}{Proposition}
\newtheorem{pr}{Proposition}
\newtheorem{remark}{Remark}
\newtheorem{re}{Remark}
\newtheorem{theorem}{Theorem}
\newtheorem{theo}{Theorem}

\def\ll{\left\lgroup}
\def\rr{\right\rgroup}

\newtheorem{Theorem}{Theorem}[section]
\newtheorem{Corollary}[Theorem]{Corollary}
\newtheorem{Proposition}[Theorem]{Proposition}
\newtheorem{Conjecture}[Theorem]{Conjecture}
\newtheorem{Lemma}[Theorem]{Lemma}
\newtheorem{Example}[Theorem]{Example}
\newtheorem{Note}[Theorem]{Note}
\newtheorem{Definition}[Theorem]{Definition}

\section{Introduction}

A large variety of two-dimensional statistical 
mechanical models 
are known to be soluble on the basis of commuting 
transfer matrices method devised in the early 1970's
by Baxter \cite{BAX}. A given lattice model is described by
its elementary Boltzmann weights $\bf{w}$ which 
can be organized as
points of $n$-dimensional 
projective space
$\mathbb{CP}^n$. This means that the weights 
can be seen
as set of $n+1$ coordinates
$[\omega_0:\omega_1:\dots:\omega_n]$
such that 
$[\lambda\omega_0:\lambda\omega_1:\dots:\lambda\omega_n]$ are identified
with $[\omega_0:\omega_1:\dots:\omega_n]$
for any non zero number $\lambda$. Let $\mathrm{T}_{\mathrm{N}}(\bf{w})$ be
the transfer matrix of a given model which depends on the size $\mathrm{N}$ 
of the lattice. Baxter's approach to integrability assumes that it is possible 
to imbed the transfer
matrix into a family of pairwise commuting operators,
\begin{equation}
\label{COM}
[\mathrm{T}_{\mathrm{N}}(\bf{w}_1)
,\mathrm{T}_{\mathrm{N}}(\bf{w}_2)]=0~~\forall~\bf{w}_1~~\mathrm{and}~~\bf{w}_2.
\end{equation}

For each value of $\mathrm{N}$ the condition (\ref{COM}) 
will lead us to a system of algebraic 
relations for the unknown 
weights $\bf{w}$. We then hope
that after certain
size $\mathrm{N}_0$ these polynomial constraints will become 
redundant in the sense that they will belong
to the ideal generated by the relations coming from previous 
lattice sizes $\mathrm{N} < \mathrm{N}_0$. We next have to be 
able to recast these finite collection of algebraic equations in the form
$\mathrm{H}_{\alpha}({\bf{w}}_1) \mathrm{G}_{\alpha}({\bf{w}}_2)-
\mathrm{G}_{\alpha}({\bf{w}}_1) \mathrm{H}_{\alpha}({\bf{w}}_2)$
where $\mathrm{H}_{\alpha}({\bf{w}})$ and $\mathrm{G}_{\alpha}({\bf{w}})$ 
are homogeneous polynomials with the same degree. 
In this
situation, the weights will be sited in 
an algebraic variety $\mathrm{X}$ 
in $\mathbb{CP}^n$ defined formally as,
\begin{equation}
\mathrm{X}=\{{\bf{w}} \in \mathbb{CP}^n~ |~ \mathrm{P}_1(\omega_0,\dots,\omega_n)=
\mathrm{P}_2(\omega_0,\dots,\omega_n)=\dots
=\mathrm{P}_k(\omega_0,\dots,\omega_n)=0 \},
\end{equation}
where
$\mathrm{P}_{\alpha}(\omega_0,\dots,\omega_n)=
\mathrm{H}_{\alpha}(\omega_0,\dots, \omega_n)-\Lambda_{\alpha} \mathrm{G}_{\alpha}(\omega_0,\dots,\omega_n)$ such that 
$\Lambda_{\alpha}$ are coupling parameters.

In fact, Baxter has introduced a finite number 
of local conditions 
which are sufficient for the commutativity of two 
distinct transfer matrices. 
This is the celebrated Yang-Baxter algebra which for 
vertex models can be expressed in terms of
product of matrices acting on three different spaces. Let us
denote by $\mathrm{L}({\bf{w}})$ the transition operator
encoding the structure of the Boltzmann weights of the given
vertex model. The transfer
matrix can be written as ordered product of such operators 
and the commutativity 
condition (\ref{COM}) is assured provided that there exists an 
invertible matrix
$\mathrm{R}({\bf{w}}_1,{\bf{w}}_2)$ satisfying the algebraic
relation,
\begin{equation}
\label{YB1}
\mathrm{R}_{12}({\bf{w}}_1,{\bf{w}}_2)
\mathrm{L}_{13}({\bf{w}}_1)
\mathrm{L}_{23}({\bf{w}}_2)=
\mathrm{L}_{23}({\bf{w}}_2)
\mathrm{L}_{13}({\bf{w}}_1)
\mathrm{R}_{12}({\bf{w}}_1,{\bf{w}}_2),
\end{equation}
where the subscript indices $ij$ denote the 
two-dimensional subspace 
in which a given operator is acting on.

At this point we emphasize that the geometrical properties 
associated to the $\mathrm{R}$-matrix can not in general 
be read directly from
that of the Boltzmann weights. From the Yang-Baxter algebra
the $\mathrm{R}$-matrix elements can be retrieved by standard
linear elimination and thus they 
define the following rational
map,
\begin{equation}
\renewcommand{\arraystretch}{1.5}
\begin{array}{ccc}
\mathrm{X} \times \mathrm{X} \subset \mathbb{CP}^n \times 
\mathbb{CP}^n &~~~ \overset{\phi(\mathrm{R})}{-~-~-\rightarrow}~ 
& \mathrm{Y} \subset \mathbb{CP}^m \\
\left[{\bf{w}}_1\right] \times \left[{\bf{w}}_2\right]
& \longmapsto &~[\phi_0({\bf{w}}_1,{\bf{w}}_2),\dots,
\phi_m({\bf{w}}_1,{\bf{w}}_2)],
 \nonumber \\
\end{array}
\end{equation}
where $m$ refers to the number of linearly 
independent $\mathrm{R}$-matrix  elements and  
$\phi_j({\bf{w}}_1,{\bf{w}}_2)$ are bi-homogeneous polynomials on two distinct sets of
weights.

The algebraic geometry of the $\mathrm{R}$-matrix 
is then described 
by $\mathrm{Y}$ which does not need 
to coincide with the geometric properties of the product
$\mathrm{X} \times \mathrm{X}$ since generically 
$\phi(\mathrm{R})$ can be a high degree map far 
from a birational equivalence\footnote{
In the special cases where $\mathrm{X}$ and 
$\mathrm{Y}$
are the same varieties the rational 
map $\phi(\mathrm{R})$ plays the role of
an addition rule among the weights
typical of algebraic groups.}. 
In order to study the geometry of 
$\mathrm{Y}$ we need to determine
its defining equations which are obtained by computing 
the implicit representation
of the image of the rational map
$\phi(\mathrm{R})$. This task is accomplished by 
eliminating the variables
${\bf{w}}_1$ and ${\bf{w}}_2$ out of the following ideal,
\begin{equation}
\label{IDEAL}
\mathrm{{\bf{I}}}=<\mathrm{P}_1({\bf{w}}_1),\dots,
\mathrm{P}_k({\bf{w}}_1);
\mathrm{P}_1({\bf{w}}_2),\dots,
\mathrm{P}_k({\bf{w}}_2);
r_0-\phi_0({\bf{w}}_1,{\bf{w}}_2),\dots,
r_m-\phi_m({\bf{w}}_1,{\bf{w}}_2)>,
\end{equation}
where $r_0,\dots,r_m$ denote the independent entries 
of the $\mathrm{R}$-matrix. 

As a result of the elimination 
procedure we will find a number of 
algebraic constraints among  the
$\mathrm{R}$-matrix elements making it possible 
to formally 
represent the variety $\mathrm{Y}$ as,
\begin{equation}
\mathrm{Y}=\{[r_0:\dots:r_m] \in \mathbb{CP}^m~ |~ \mathrm{Q}_1(r_0,\dots,r_m)=
\mathrm{Q}_2(r_0,\dots,r_m)=\dots
=\mathrm{Q}_l(r_0,\dots,r_m)=0 \},
\end{equation}
where $\mathrm{Q}_{\alpha}(r_0,\dots,r_m)$ are yet another 
family of homogenous polynomials. 

In principle, the explicit expressions for 
$\mathrm{Q}_{\alpha}(r_0,\dots,r_m)$
can be obtained by using an alternative
representation for the ideal $\mathrm{{\bf{I}}}$ denominated  
Groebner bases \cite{COX}. However, in practice
this task is algorithmically involved 
depending much on the 
complexity of the polynomials
defining both $\mathrm{X}$ and the rational map
$\phi(\mathrm{R})$. For an example in the 
case of the vertex model  
associated to the Hubbard chain we refer to \cite{MA}.

Having at hand a solution of the 
Yang-Baxter equation 
it is of interest to uncover 
the geometric
content of both varieties $\mathrm{X}$ and $\mathrm{Y}$. This is
specially relevant in situations where the 
elements of the $\mathrm{R}$-matrix 
are not all expressed in terms of rational functions. The presence of multiple 
coverings such as
square roots
terms could hide the actual 
geometric content underlying
the Yang-Baxter solution. In this paper we investigate this issue
for a
$\mathrm{R}$-matrix based on a deformation of the
centrally extended $sl(2|2)$ superalgebra found 
by Beisert and Koroteev within
the quantum group
machinery \cite{BEKO}. 
For generic gauge
we show that the Boltzmann weights sit on a surface ruled by an
elliptic curve which has a degree two isogeny with the 
genus one parameterization devised by Beisert and Koroteev. It turns out that
the suitable symmetric gauge choice made 
in \cite{BEKO} does not cut
the ruled surface on its $\mathbb{CP}^1$ fibre and in this case
the Boltzmann weights lie on a curve of genus five. The geometric
properties of the $\mathrm{R}$-matrix for generic gauge 
are argued to be governed
by the product variety $\mathbb{CP}^1 \times \mathbb{CP}^1 \times \mathrm{A}$ where
$\mathrm{A}$ is an Abelian surface. However, for
the symmetric calibration,  we present strong evidences that the 
surface is of general type sitting 
on the Severi line \cite{SE}, that is,  
the canonical class $\mathcal{K}_{\mathrm{S}}$ and the Euler-Poincar\'e characteristic
$\chi({\mathrm{S}})$ of the surface satisfy the
relation 
$\mathcal{K}^2_{\mathrm{S}}=4 \chi({\mathrm{S}})$.
These results generalize in a substantial way the recent
work \cite{MA} associated to the specific undeformed case.

We have organized this paper as follows. In next section we 
derive an alternative 
representation for the $\mathrm{R}$-matrix such that the matrix
elements are rational functions of certain elementary weights
sited on a degree six surface. This provides the basics to
investigate the geometrical properties associated to  both
the Boltzmann weights and the $\mathrm{R}$-matrix performed in
sections 3. In section 4 we discuss the geometrical content
in the interesting case of a symmetric gauge choice. Our 
conclusions are in section 5 and 
in three appendices we summarize some
technical details omitted in the main text.

\section{The $q$-deformed $\mathrm{R}$-matrix}

We start recalling that the four-dimensional 
representation of the quantum 
deformation of the extended $sl(2|2)$ superalgebra has been 
parametrized in terms of three 
variables denoted by $x^{+},x^{-}$ and $\gamma$. 
The latter plays the 
role of a free gauge parameter while $x^{\pm}$ are 
required to fulfill the following elliptic curve \cite{BEKO},
\begin{equation}
\label{E1}
\mathrm{E}_1=\frac{x_{+}}{q}+\frac{q}{x_{+}} -qx_{-}-\frac{1}{qx_{-}}+\IM g(q-1/q)(\frac{x_{+}}{qx_{-}}-\frac{qx_{-}}{x_{+}})-\frac{\IM}{g},
\end{equation}
where $q$ denotes the deformation parameter and $g$ is a coupling constant. 

The intertwining operator encoding the graded structure of such fundamental
representation has been originally constructed by Beisert and Koroteev \cite{BEKO}. For the
purposes of this paper it is enough to the consider the related 
$\mathrm{R}$-matrix satisfying the standard Yang-Baxter
equation, namely
\begin{equation}
\label{YB}
\mathrm{R}_{12}({\bf{w}}_1,{\bf{w}}_2)
\mathrm{R}_{13}({\bf{w}}_1,{\bf{w}}_3)
\mathrm{R}_{23}({\bf{w}}_2,{\bf{w}}_3)=
\mathrm{R}_{23}({\bf{w}}_2,{\bf{w}}_3)
\mathrm{R}_{13}({\bf{w}}_2,{\bf{w}}_3)
\mathrm{R}_{12}({\bf{w}}_1,{\bf{w}}_2).
\end{equation}

Following the original work \cite{BEKO} we can represent the operator $\mathrm{R}$
by the following matrix,
\begin{equation}
\label{RMABK}
\mathrm{R}=\left(
\begin{array}{cccc|cccc|cccc|cccc}
\mathcal{A}& 0& 0& 0& 0& 0& 0& 0& 0& 0& 0& 0& 0& 0& 0& 0 \\
0& \mathcal{B}& 0& 0& \mathcal{C}& 0& 0& 0& 0& 0& 0& 0& 0& 0& 0& 0 \\
0& 0& \mathcal{B}& 0& 0& 0& 0& 0& \mathcal{C}& 0& 0& 0& 0& 0& 0& 0 \\
0& 0& 0& \mathcal{F}& 0& 0& \frac{\mathcal{D}}{\delta}& 0& 0& \frac{-q\mathcal{D}}{\delta}& 0& 0& {\mathcal{A}}-q\mathcal{F}& 0& 
0& 0 \\ \hline
0& \overline{\mathcal{C}}& 0& 0& \overline{\mathcal{B}} & 0& 0& 0& 0& 
      0& 0& 0& 0& 0& 0& 0 \\
0& 0& 0& 0& 0& 1 & 0& 
      0& 0& 0& 0& 0& 0& 0& 0& 0 \\
0& 0& 0& -\delta q \overline{\mathcal{D}} & 0& 0& 
      \mathcal{G}& 0& 0& 1-q\mathcal{G} & 0& 0& \delta {q}^2 \overline{\mathcal{D}} & 0& 0& 0 \\
     0& 0& 0& 0& 0& 0& 0& \overline{\mathcal{B}} & 0& 0& 0& 0& 0& 
      \overline{\mathcal{C}} & 0& 0 \\ \hline
0& 0& \overline{\mathcal{C}} & 0& 0& 0& 0& 0& 
      \overline{\mathcal{B}} & 0& 0& 0& 0& 0& 0& 0 \\
     0& 0& 0& \delta \overline{\mathcal{D}} & 0& 0& 1-\frac{\mathcal{G}}{q} & 0& 0& \mathcal{G} & 0& 
      0& -\delta {q} \overline{\mathcal{D}} & 0& 0& 0 \\
0& 0& 0& 0& 0& 0& 0& 0& 0& 0& 
      1 & 0& 0& 0& 0& 0 \\ 
     0& 0& 0& 0& 0& 0& 0& 0& 0& 0& 0& \overline{\mathcal{B}} & 0& 0& 
      \overline{\mathcal{C}} & 0 \\ \hline
0& 0& 0& {\mathcal{A}} -\frac{{\mathcal{F}}}{q}& 0& 0& \frac{{-\mathcal{D}}}{\delta q} & 0& 0& \frac{\mathcal{D}}{\delta} & 0& 
      0& \mathcal{F} & 0& 0& 0 \\
0& 0& 0& 0& 0& 0& 0& \mathcal{C} & 0& 0& 0& 0& 0& 
      \mathcal{B} & 0& 0 \\
0& 0& 0& 0& 0& 0& 0& 0& 0& 0& 0& \mathcal{C} & 0& 0& 
      \mathcal{B} & 0 \\
0& 0& 0& 0& 0& 0& 0& 0& 0& 0& 0& 0& 0& 0& 0& \mathcal{A} \\
\end{array}
\right),
\end{equation}
where $\delta$ is a free twist parameter. 

Defining the auxiliary parameter 
$\xi=\IM g (q-1/q)$ the matrix elements can be expressed as,
\begin{eqnarray}
\label{WEIGB}
&& \mathcal{A}=\frac{(x_1^{-}-x_2^{+})\sqrt{(\xi+x_1^{+})(\xi+x_2^{-})}}{(x_2^{-}-x_1^{+})\sqrt{(\xi+x_2^{+})(\xi+x_1^{-})}} 
,~~\mathcal{B}=\frac{(x_1^{+}-x_2^{+})\sqrt{\xi+x_2^{-}}}{\sqrt{q}(x_1^{+}-x_2^{-})\sqrt{\xi+x_2^{+}}}, \nonumber \\ \nonumber \\
&& \overline{\mathcal{B}}=\frac{\sqrt{q}(x_1^{-}-x_2^{-})\sqrt{\xi+x_1^{+}}}{(x_1^{+}-x_2^{-})\sqrt{\xi+x_1^{-}}}
,~~\mathcal{C}=\frac{\gamma_2(x_1^{-}-x_1^{+})\sqrt{(\xi+x_1^{+})(\xi+x_2^{-})}}{\gamma_1(x_2^{-}-x_1^{+})\sqrt{(\xi+x_2^{+})(\xi+x_1^{-})}},
\nonumber \\ \nonumber \\
&& \overline{\mathcal{C}}=\frac{\gamma_1(x_2^{-}-x_2^{+})}{\gamma_2(x_2^{-}-x_1^{+})}
,~~\mathcal{D}=\frac{(x_1^{-}-x_1^{+})(x_2^{-}-x_2^{+})(x_2^{+}-x_1^{+})\sqrt{\xi+x_2^{-}}}{\gamma_1\gamma_2(x_2^{-}-x_1^{+})\sqrt{\xi+x_2^{+}}[1-\xi(x_1^{-}+x_2^{-})-x_1^{-}x_2^{-}]}, \nonumber \\ \nonumber \\
&& \overline{\mathcal{D}}=\frac{\gamma_1\gamma_2(1+\xi^2)(x_2^{+}-x_1^{+})(\xi+x_2^{-})\sqrt{\xi+x_1^{-}}}{q^3(x_2^{-}-x_1^{+})(\xi+x_2^{+})\sqrt{\xi+x_1^{+}}[1-\xi(x_1^{-}+x_2^{-})-x_1^{-}x_2^{-}]}, \nonumber \\ \nonumber \\
&& \mathcal{F}=\frac{(x_1^{+}-x_2^{+})\sqrt{(\xi+x_1^{-})(\xi+x_2^{-})}[1-\xi(x_1^{+}+x_2^{-})-x_1^{+}x_2^{-}]}{q(x_2^{-}-x_1^{+})\sqrt{(\xi+x_1^{+})(\xi+x_2^{+})}[1-\xi(x_1^{-}+x_2^{-})-x_1^{-}x_2^{-}]}, \nonumber \\ \nonumber \\
&& \mathcal{G}=\frac{(\xi+x_2^{-})(x_2^{+}-x_1^{+})[1-\xi(x_1^{-}+x_2^{+})-x_1^{-}x_2^{+}]}{q(\xi+x_2^{+})(x_2^{-}-x_1^{+})(1-\xi(x_1^{-}+x_2^{-})-x_1^{-}x_2^{-}]},
\end{eqnarray}
where the subscript index $j$ means distinct points $x_j^{\pm}$ on the
curve (\ref{E1}).

In this representation we see that many of the 
matrix elements (\ref{WEIGB}) 
contain 
square root terms. This fact hides the actual geometric content associated to such solution
of the Yang-Baxter equation (\ref{YB}).
The ideal situation is to have 
all the $\mathrm{R}$-matrix elements 
written only in terms
of ratios of bi-homogeneous polynomials and a
systematic way to 
uncover such algebraic structure is as follows. We first expand one of the $\mathrm{R}$-matrix 
rapidities pair around a generic point belonging to 
the curve (\ref{E1}). The next step is to identify
some of the expanded entries of the $\mathrm{R}$-matrix with the coordinates $[x:y:z:w]$
of a three-dimensional projective space. This leads to constraints which need to be solved for
the variables $x^{\pm},\gamma$ and afterwards we should verify that 
all the matrix elements
are indeed rational functions on the ring
$\mathbb{C}[x,y,z,w]$. It turns out that one possible reference point is,
\begin{equation}
x^{+}=-\xi +\epsilon,~~~x^{-}=\frac{1+\xi^2}{q^2}\frac{1}{\epsilon},~~~\gamma=\frac{1}{q^{1/4}}+\epsilon,
\end{equation}
where $\epsilon$ is an expansion variable.

We now expand the second set of rapidities of 
the $\mathrm{R}$-matrix (\ref{RMABK},\ref{WEIGB}) around 
the above point and search
for the simplest ratios among the expanded matrix entries. We find that these
are given by the amplitudes $\mathcal{A}/\mathcal{C}$,
$\mathcal{B}/\mathcal{C}$ and
$\overline{\mathcal{C}}/\mathcal{C}$ which in the limit 
$\epsilon \rightarrow 0$ gives, respectively,
the identification,
\begin{equation}
\frac{x}{w}=\frac{q^{1/4} \gamma (\xi+x^{-})}{x^{-}-x^{+}},~~
\frac{y}{w}=\frac{\gamma \sqrt{(\xi+x^{-})(\xi+x^{+})}}{q^{1/4}(x^{+}-x^{-})},~~
\frac{z}{w}=\frac{\gamma^2 \sqrt{(1+\xi^2)(\xi+x^{-})}}{q^{1/2}(x^{-}-x^{+})\sqrt{\xi+x^{+}}}.
\end{equation}

In order to solve the above constraints for the 
variables $x^{\pm}$ and $\gamma$ 
we consider the relations coming from the ratios $x^2/y^2$ and $y/z$. This
provides three linear equations which are easily solved and the final result is,
\begin{equation}
\label{CHAN}
x_{+}=-\xi-\frac{\sqrt{1+\xi^2}}{\sqrt{q}}\left(\frac{y}{x}\right)\frac{(x^2-qy^2)}{zw},~~
x_{-}=-\xi-\frac{\sqrt{1+\xi^2}}{q^{3/2}}\left(\frac{x}{y}\right)\frac{(x^2-qy^2)}{zw},~~
\gamma=\frac{x^2-qy^2}{q^{1/4}xw}.
\end{equation}

By using the above relations we have checked that matrix 
elements of the $\mathrm{R}$-matrix
expansion are indeed expressed in terms of ratios 
of polynomials on $\mathbb{C}[x,y,z,w]$. Note that relations (\ref{CHAN}) define
a two-to-one mapping on the affine space $w=1$ and therefore the geometrical properties 
are not fully captured by the algebraic geometry description
on the variables $x^{\pm}$ and $\gamma$. The proper geometric content of the Boltzmann
weights should be uncovered 
from the polynomial 
constraining the homogeneous variables $x,y,z,w$ which is
obtained by substituting Eq.(\ref{CHAN}) into the original elliptic curve (\ref{E1}).
After some cumbersome simplifications we find that such algebraic surface is defined by, 
\begin{equation}
\label{SUP}
\mathrm{S}=(x^2-\frac{y^2}{q})(x^2-qy^2)^2-\mathrm{U}xyzw(x^2-qy^2)-w^2z^2(x^2-q^3y^2),
\end{equation}
where the Hubbard like coupling $\mathrm{U}$ is,
\begin{equation}
\mathrm{U}=\frac{\sqrt{q}\left[1-2g^2(q-1/q)^2\right]}{g\sqrt{g^2(q-1/q)^2-1}}.
\end{equation}

Finally, we present the expression for the $\mathrm{R}$-matrix 
in terms of the surface $\mathrm{S}$ variables. As before its basic
matrix structure is given by,
\begin{equation}
\mathrm{R}=\left(
\begin{array}{cccc|cccc|cccc|cccc}
\bf{a}& 0& 0& 0& 0& 0& 0& 0& 0& 0& 0& 0& 0& 0& 0& 0 \\
0& \bf{b}& 0& 0& \bf{c}& 0& 0& 0& 0& 0& 0& 0& 0& 0& 0& 0 \\
0& 0& \bf{b}& 0& 0& 0& 0& 0& \bf{c}& 0& 0& 0& 0& 0& 0& 0 \\
0& 0& 0& \bf{f}& 0& 0& \frac{\bf{d}}{\delta_1} & 0& 0& -\frac{q\bf{d}}{\delta_1}& 0& 0& {\bf{a}}-q\bf{f}& 0& 
0& 0 \\ \hline
0& \overline{\bf{c}}& 0& 0& \overline{\bf{b}} & 0& 0& 0& 0& 
      0& 0& 0& 0& 0& 0& 0 \\
0& 0& 0& 0& 0& \bf{g} & 0& 
      0& 0& 0& 0& 0& 0& 0& 0& 0 \\
0& 0& 0& -q\delta_1\overline{\bf{d}} & 0& 0& 
      \overline{\bf{g}}& 0& 0& {\bf{g}}-q\overline{\bf{g}} & 0& 0& q^2\delta_1\overline{\bf{d}} & 0& 0& 0 \\
     0& 0& 0& 0& 0& 0& 0& \overline{\bf{b}} & 0& 0& 0& 0& 0& 
      \overline{\bf{c}} & 0& 0 \\ \hline
0& 0& \overline{\bf{c}} & 0& 0& 0& 0& 0& 
      \overline{\bf{b}} & 0& 0& 0& 0& 0& 0& 0 \\ 
     0& 0& 0& \delta_1 \overline{\bf{d}} & 0& 0& {\bf{g}}-\frac{\overline{{\bf{g}}}}{q} & 0& 0& \overline{\bf{g}} & 0 & 
      0& -q\delta_1 \overline{\bf{d}} & 0& 0& 0 \\ 
0& 0& 0& 0& 0& 0& 0& 0& 0& 0& 
      \bf{g} & 0& 0& 0& 0& 0 \\ 
     0& 0& 0& 0& 0& 0& 0& 0& 0& 0& 0& \overline{\bf{b}} & 0& 0& 
      \overline{\bf{c}} & 0 \\ \hline
0& 0& 0& {\bf{a}} -\frac{{\bf{f}}}{q\delta_1}& 0& 0& -\frac{\bf{d}}{q\delta_1} & 0& 0& \frac{\bf{d}}{\delta_1} & 0& 
      0& \bf{f} & 0& 0& 0 \\
0& 0& 0& 0& 0& 0& 0& \bf{c} & 0& 0& 0& 0& 0& 
      \bf{b} & 0& 0 \\
0& 0& 0& 0& 0& 0& 0& 0& 0& 0& 0& \bf{c} & 0& 0& 
      \bf{b} & 0 \\
0& 0& 0& 0& 0& 0& 0& 0& 0& 0& 0& 0& 0& 0& 0& \bf{a} \\
\end{array}
\right) \nonumber \\
\end{equation}
where the twist relation is $\delta_1=-\delta/q$.

The matrix elements are obtained by using the 
mapping (\ref{CHAN}) on the previous amplitudes
given by Eq.(\ref{WEIGB}). After some algebra and up to 
an overall normalization we obtain,
\begin{eqnarray}
\label{RMAG}
&&\frac{{\bf{a}}}{{\bf{c}}}=\frac{{\bf{x}}_1{\bf{x}}_2}{\theta({\bf{x}}_2,{\bf{y}}_2)}-q\frac{\bf{z}_1}{\bf{z}_2}\frac{{\bf{y}}_1{\bf{y}}_2}{\theta({\bf{x}}_1,{\bf{y}}_1)},~~
\frac{{\bf{b}}}{{\bf{c}}}=\frac{{\bf{y}}_1{\bf{x}}_2}{\theta({\bf{x}}_2,{\bf{y}}_2)}-\frac{\bf{z}_1}{\bf{z}_2}\frac{{\bf{x}}_1{\bf{y}}_2}{\theta({\bf{x}}_1,{\bf{y}}_1)},~~\frac{\overline{\bf{c}}}{{\bf{c}}}= \frac{{\bf{z}}_1}{{\bf{z}}_2}, \nonumber \\ \nonumber \\
&& \frac{\overline{\bf{b}}}{{\bf{c}}}=q\frac{{\bf{x}}_1{\bf{y}}_2}{\theta({\bf{x}}_2,{\bf{y}}_2)}-q\frac{\bf{z}_1}{\bf{z}_2}\frac{{\bf{y}}_1{\bf{x}}_2}{\theta({\bf{x}}_1,{\bf{y}}_1)},~~
\frac{{\bf{g}}}{{\bf{c}}}=\frac{\bf{z}_1}{\bf{z}_2}\frac{{\bf{x}}_1{\bf{x}}_2}{\theta({\bf{x}}_1,{\bf{y}}_1)}-q\frac{{\bf{y}}_1{\bf{y}}_2}{\theta({\bf{x}}_2,{\bf{y}}_2)}, \nonumber \\ \nonumber \\
&& \frac{{\bf{d}}}{{\bf{c}}}=\frac{{\bf{x}}_1{\bf{y}}_1\theta({\bf{x}}_1,{\bf{y}}_1)({\bf{x}}_2^2-q^3{\bf{y}}_2^2)-\frac{\bf{z}_1}{\bf{z}_2}{\bf{x}}_2{\bf{y}}_2\theta({\bf{x}}_2,{\bf{y}}_2)({\bf{x}}_1^2-q^3{\bf{y}}_1^2)}{\theta({\bf{x}}_1,{\bf{y}}_1)\theta({\bf{x}}_2,{\bf{y}}_2)({\bf{x}}_1^2{\bf{x}}_2^2-q^2{\bf{y}}_1^2{\bf{y}}_2^2)}
,~~\frac{\overline{\bf{d}}}{{\bf{c}}}= {\bf{z}}_1{\bf{z}}_2 
\frac{{\bf{d}}}{{\bf{c}}},
 \nonumber \\ \nonumber \\
&&\frac{{\bf{f}}}{{\bf{c}}}=\frac{{\bf{x}}_1{\bf{y}}_1\left[{\bf{x}}_2{\bf{y}}_1\theta({\bf{x}}_1,{\bf{y}}_1)-\frac{\bf{z}_1}{\bf{z}_2}
{\bf{x}}_1{\bf{y}}_2\theta({\bf{x}}_2,{\bf{y}}_2)\right]}{\theta({\bf{x}}_1,{\bf{y}}_1)({\bf{x}}_1^2{\bf{x}}_2^2-q^2{\bf{y}}_1^2{\bf{y}}_2^2)}
+\frac{q^2{\bf{x}}_2{\bf{y}}_2\left[{\bf{x}}_1{\bf{y}}_2\theta({\bf{x}}_1,{\bf{y}}_1)-\frac{\bf{z}_1}{\bf{z}_2}{\bf{x}}_2{\bf{y}}_1\theta({\bf{x}}_2,{\bf{y}}_2)\right]}{\theta({\bf{x}}_2,{\bf{y}}_2)({\bf{x}}_1^2{\bf{x}}_2^2-q^2{\bf{y}}_1^2{\bf{y}}_2^2)}, \nonumber \\ \nonumber \\
&&\frac{\overline{\bf{g}}}{{\bf{c}}}=
\frac{\left[q^2{\bf{z}_1}{\bf{z}_2}{\bf{x}}_2{\bf{y}}_1-{\bf{x}}_1{\bf{y}}_2\theta({\bf{x}}_1,{\bf{y}}_1)\theta({\bf{x}}_2,{\bf{y}}_2)\right]}
{\theta({\bf{x}}_1,{\bf{y}}_1)\theta({\bf{x}}_2,{\bf{y}}_2)}
\frac{{\bf{d}}}{{\bf{c}}},
\end{eqnarray}
where 
$\theta({\bf{x}},{\bf{y}})={\bf{x}}^2-q{\bf{y}}^2$ and
the bold letters refer to the affine coordinates
${\bf{x}}=x/w$,
${\bf{y}}=y/w$,
and ${\bf{z}}=z/w$.

As expected the $\mathrm{R}$-matrix are expressed solely in terms of ratios of bi-homogeneous
polynomials on the coordinates ${\bf{x}}_j$, ${\bf{y}}_j$ and ${\bf{z}}_j$. 
At this point we have gathered the basic ingredients to study the geometrical properties
of both the Boltzmann weights and the $\mathrm{R}$-matrix.

\section{Algebraic geometry for arbitrary gauge}

We start by analyzing the geometrical properties of the Boltzmann weights associated 
to the sextic surface (\ref{SUP}).
This surface contains one-dimensional singularities
which in principle should be resolved by means of
birational morphisms. A partial desingularization is performed eliminating the product term
$zw$ by means of standard quadrature.  This makes it possible to decrease the degree of
the surface polynomial and as a result we have the following map,
\begin{equation}
\renewcommand{\arraystretch}{1.5}
\begin{array}{ccc}
\mathrm{S} \subset \mathbb{CP}^3  
&~~~ \overset{\phi}{-~-~-\rightarrow}~~~ 
& \widetilde{\mathrm{S}} \subset\mathbb{CP}^3 \\
$[x:y:z:w]$
& \longmapsto &~[\frac{\phi_1}{\phi_2}:w:x:y],
\end{array}
\end{equation}
where the polynomials
$\phi_1=\IM \sqrt{q}[\mathrm{U}(x^2-qy^2)xy+2(x^2-q^3y^2)zw]$ and 
$\phi_2=(x^2-qy^2)w$ while $\widetilde{\mathrm{S}}$ is a degree four surface
defined by,
\begin{equation}
\widetilde{\mathrm{S}}=x_0^2x_1^2+4qx_2^4-(4-q\mathrm{U}^2+4q^4)x_2^2x_3^2+4q^3x_3^4.
\end{equation}

The above map defines a birational equivalence 
since it is invertible
away from the singular locus 
of the surface $\mathrm{S}$. The
inverse map is given by,
\begin{equation}
\renewcommand{\arraystretch}{1.5}
\begin{array}{ccc}
\widetilde{\mathrm{S}} \subset \mathbb{CP}^3  
&~~~ \overset{\phi^{-1}}{-~-~-\rightarrow}~~~ 
& \mathrm{S} \subset\mathbb{CP}^3 \\
$[$x_0$:$x_1$:$x_2$:$x_3$]$
& \longmapsto &~[x_2:x_3:\frac{\psi_2}{\psi_1}:x_1],
\end{array}
\end{equation}
where $\psi_1=2\IM\sqrt{q}x_1(x_2^2 -q^3x_3^2)$ and
$\psi_2=(x_0x_1 - \IM\sqrt{q}\mathrm{U}x_2x_3)(x_2^2 - qx_3^2)$.

The next step is to observe that out of $\widetilde{\mathrm{S}}$ 
one can define a surjective projection to an elliptic curve,
\begin{equation}
\renewcommand{\arraystretch}{1.5}
\begin{array}{ccc}
\widetilde{\mathrm{S}} \subset \mathbb{CP}^3  
&~~~ \overset{\pi} {\longrightarrow}~~~ 
& \mathrm{E}_2 \subset\mathbb{CP}^2, 
\end{array}
\end{equation}
such that the fibre $\pi^{-1}$ at every
point on $\mathrm{E}_2$
is isomorphic to $\mathbb{CP}^1$.
The affine form of $\mathrm{E}_2$ 
is that of a Jacobi quartic, namely
\begin{equation}
\mathrm{E}_2=y_1^2+4q-(4-q\mathrm{U}^2+4q^4)y_2^2+4q^3y_2^4.
\end{equation}

Putting all these results together we conclude that
$\mathrm{S}$ is in fact a surface ruled by a genus one curve, 
that is $\mathrm{S} \cong \mathbb{CP}^1 \times \mathrm{E}_2$.
In order to provide a concrete representation for the surface
$\mathrm{S}$ variables we associate the coordinate $t$ to the affine part 
of its $\mathbb{CP}^1$ subspace.
Taking into account the form of the inverse map $\phi^{-1}$ 
we can express the ratios of the 
surface $\mathrm{S}$ variables
as follows,
\begin{equation}
\frac{x}{w}=t,~~\frac{y}{w}= t y_2,~~\frac{z}{w}=t^2
\frac{(y_1 - \IM\sqrt{q}\mathrm{U}y_2)(1 - qy_2^2)}{
2\IM\sqrt{q}(1 -q^3y_2^2)}.
\end{equation}
whose uniformization depends only 
on the curve $\mathrm{E}_2$. In the appendix $A$
we present one such uniformization for 
the variables $y_1$ and $y_2$.

We next remark that the elliptic curves 
$\mathrm{E}_1$ and $\mathrm{E}_2$ are 
not isomorphic 
but only twofold
isogenous. Note that the degree of the isogeny is 
in accordance with 
the map (\ref{CHAN}) degree. A simple way to see this 
fact is through the comparison of
their $\mathrm{J}$-invariants since they will 
fix two points on 
the genus one curve moduli
space \cite{SIL}. In terms of the deformation
parameter $q$ 
and the coupling $\mathrm{U}$ the expressions 
of the corresponding
$\mathrm{J}$-invariants are,
\begin{eqnarray}
\label{JIN}
\mathrm{J}(\mathrm{E}_1)&=&\frac{\left(16-8q\mathrm{U}^2+q^2\mathrm{U}^4-16q^4-8q^5\mathrm{U}^2+16q^8\right)^3}{q^8(4-q\mathrm{U}^2+4q^4+8q^2)(4-q\mathrm{U}^2+4q^4-8q^2)} \nonumber \\ \nonumber \\
\mathrm{J}(\mathrm{E}_2)&=&\frac{\left(16-8q\mathrm{U}^2+q^2\mathrm{U}^4+224q^4-8q^5\mathrm{U}^2+16q^8\right)^3}{q^4(4-q\mathrm{U}^2+4q^4+8q^2)^2(4-q\mathrm{U}^2+4q^4-8q^2)^2}.
\end{eqnarray}

Because the $\mathrm{J}$-invariants are not the same the two curves can not 
be isomorphic for arbitrary
values of the parameters $q$ and $\mathrm{U}$. They are however related by an isogeny 
of degree two and this feature
can be verified with the help of the 
respective modular polynomial. This is a symmetric two variable
polynomial with suitable coefficients and its 
explicit expression is,
\begin{eqnarray}
\label{MOD}
\Phi_2[x,y]&=&x^3 +y^3- x^2y^2 + 1488xy(x+y) - 162000(x^2+y^2) 
+40773375xy \nonumber \\ 
&+& 8748000000(x+y)
-157464000000000.
\end{eqnarray}

Two elliptic curves are said to be twofold isogenous 
provided the valuation of the modular polynomial (\ref{MOD})
at distinct $\mathrm{J}$-invariants
is zero. Taking into account the expressions (\ref{JIN}) we find 
indeed that $\Phi_2\left[\mathrm{J}(\mathrm{E}_1),\mathrm{J}(\mathrm{E}_2)\right]=0$.

\subsection{$\mathrm{R}$-matrix geometry}

Here we shall investigate
the geometrical properties 
associated to the $\mathrm{R}$-matrix. As discussed in the 
introduction the first step is to obtain the defining equations 
for the corresponding variety $\mathrm{Y}$. This requires the 
elimination of the weights 
${\bf{x}}_j,
{\bf{y}}_j$ and 
${\bf{z}}_j$ from the ideal (\ref{IDEAL}) built out of the 
bi-homogenous polynomials
(\ref{RMAG}). The technicalities concerning this task are 
somehow similar to the elimination problem
solved recently for the undeformed case \cite{MA}.
Making the due adaptations  
we find that $\mathrm{Y} \in \mathbb{CP}^9$ 
is described as the intersection
of four quadrics and one quartic polynomial,
\begin{equation}
\label{YY}
\mathrm{Y}=\{ 
({\bf{a}}:{\bf{b}}:\overline{{\bf{b}}}:{\bf{c}}:\overline{{\bf{c}}}:{\bf{d}}:
\overline{{\bf{d}}}:{\bf{f}}:{\bf{g}}:\overline{{\bf{g}}}) 
\in \mathbb{CP}^9~|~\mathrm{Q}_1=\mathrm{Q}_2=\mathrm{Q}_3=\mathrm{Q}_4=\mathrm{Q}_5=0\},
\end{equation}
where the expressions for the polynomials $\mathrm{Q}_j$ are,
\begin{eqnarray}
\label{INTER}
&&\mathrm{Q}_1={\bf{b}}\overline{\bf{b}}+{\bf{a}}{\bf{g}}-{\bf{c}}\overline{{\bf{c}}},~~
\mathrm{Q}_2=\overline{\bf{b}}{\bf{b}}+{\bf{f}}\overline{\bf{g}}+q{\bf{d}}\overline{{\bf{d}}}, \nonumber \\
&&\mathrm{Q}_3={\bf{g}}{\bf{f}}+{\bf{a}}\overline{\bf{g}}+(q+1/q){\bf{b}}\overline{\bf{b}},~~
\mathrm{Q}_4={\bf{a}}{\bf{f}}+{\bf{g}}\overline{\bf{g}}-({\bf{b}}^2+\overline{\bf{b}}^2), \nonumber \\
&&\mathrm{Q}_5=\left[(\overline{{\bf{g}}}-q{\bf{g}})({\bf{g}}-q\overline{\bf{g}})
-({\bf{f}}-q{\bf{a}})({\bf{a}}-q{\bf{f}})\right]^2
-q^2\mathrm{U}^2{\bf{c}}\overline{{\bf{c}}}{\bf{d}}\overline{{\bf{d}}}.
\end{eqnarray}

It turns out that $\mathrm{Y}$ is a
four-dimensional 
complex algebraic variety such that part of its geometry is dominated by
a two-dimensional projective space. This fact can be understood by 
noticing that in the subspace
$\mathbb{C}[{\bf{c}},\overline{{\bf{c}}} 
,{\bf{d}},\overline{{\bf{d}}}]$ the polynomials contain only monomials of the form 
${\bf{c}}\overline{{\bf{c}}}$ and 
${\bf{d}}\overline{{\bf{d}}}$. They can be linearly eliminated with the help of
the first two quadrics and the remaining equations become defined 
on the complementary ring $\mathbb{C}[{\bf{a}},{\bf{b}},\overline{{\bf{b}}},
{\bf{f}},{\bf{g}},\overline{{\bf{g}}}]$. This means that $\mathrm{Y}$
is birational to the product
$\mathbb{CP}^1 \times \mathbb{CP}^1 \times \mathrm{A}$ where $\mathrm{A}$
is a surface defined by three polynomials, namely
\begin{equation}
\mathrm{A}=\{ 
({\bf{a}}:{\bf{b}}:\overline{{\bf{b}}}
:{\bf{f}}:{\bf{g}}:\overline{{\bf{g}}}) 
\in \mathbb{CP}^5~|~\mathrm{Q}_3=\mathrm{Q}_4=\widetilde{\mathrm{Q}}_5=0\},
\end{equation}
where the expression for 
$\widetilde{\mathrm{Q}}_5$ is,
\begin{equation}
\label{Q5}
\widetilde{\mathrm{Q}}_5=\left[(\overline{{\bf{g}}}-q{\bf{g}})({\bf{g}}-q\overline{\bf{g}})
-({\bf{f}}-q{\bf{a}})({\bf{a}}-q{\bf{f}})\right]^2
+q\mathrm{U}^2
({\bf{b}}\overline{\bf{b}}+{\bf{a}}{\bf{g}})
(\overline{\bf{b}}{\bf{b}}+{\bf{f}}\overline{\bf{g}}).
\end{equation}

At this point it remains to understand the 
geometric properties
of the surface $\mathrm{A}$. This investigation involves
some technical steps
summarized in Appendix B and in what follows we present 
the main conclusions. We first observe that by extracting
the monomial ${\bf{b}}\overline{\bf{b}}$ from third quadric
the polynomial $\widetilde{\mathrm{Q}}_5$
defines a surface on the ring subspace
$\mathbb{C}[{\bf{a}},
{\bf{f}},{\bf{g}},\overline{{\bf{g}}}]$.
The analysis of
the geometry of such surface reveals that it is birational 
to a surface ruled by an elliptic curve 
isomorphic to $\mathrm{E}_2$. The next step consists on the study of 
the normalization of the curve defined by the
remaining  polynomial $\mathrm{Q}_4$.
As a result we obtain that it is another genus one curve whose 
Weierstrass form is,
\begin{eqnarray}
\mathrm{E}_3 &=& y^3-x^3
+\frac{\left[16 +8q(120q -\mathrm{U}^2)(1+q^4)+q^2\mathrm{U}^4-240q^3\mathrm{U}^2+2144q^4+16q^8 \right]}{48}x \nonumber \\
&+&\frac{\left[4+4q^4+24q^2-q\mathrm{U}^2\right]\left[16-8q(264q+\mathrm{U}^2)(1+q^4)+q^2\mathrm{U}^4+528q^3\mathrm{U}^2-4000q^4+16q^8\right]}{864}. \nonumber \\
\end{eqnarray}

Collecting the above information together we are able to 
conclude that $\mathrm{A}$ is 
birational to an Abelian surface 
determined by the product of two elliptic curves,
\begin{equation}
\mathrm{A} \cong \bar{\mathrm{E}}_2 \times \bar{\mathrm{E}}_3.
\end{equation}

We would like to close this section emphasizing that 
our conclusions for the 
geometric content are valid
for generic points of the two-dimensional space of
the couplings $q$ and $\mathrm{U}$. When these parameters
are constrained to certain subspaces the respective varieties
become reducible and this changes the geometrical properties.
Potential geometric degenerations in the context of 
elliptic curves occur at the singularities of their
$\mathrm{J}$-invariants. Inspecting
Eqs.(\ref{JIN}) we see that this happens when,
\begin{equation}
\label{SUBM}
q \mathrm{U}^2-4(q^2 +\varepsilon)^2=0~~\mathrm{with}~~\varepsilon=\pm 1.
\end{equation}

In fact, under the condition (\ref{SUBM}) the sextic
surface (\ref{SUP}) becomes
reducible in terms of the product of two 
cubic surfaces. The expressions 
of the irreducible components $\overline{\mathrm{S}}_{\pm}$ are,
\begin{equation}
\overline{\mathrm{S}}_{\pm}=x^3 \pm x^2y/\sqrt{q}-qxy^2 \mp \sqrt{q}y^3 \pm \varepsilon xzw-q^{3/2}yzw,
\end{equation}
and now the 
weights sit on
a rational manifold since irreducible 
cubic surfaces are known
to be birationally isomorphic to $\mathbb{CP}^2$ \cite{BU}.

Similar scenario is also expected for the geometry 
underlying the $\mathrm{R}$-matrix and one direct way to
see such decomposition is through
a three-dimensional embedding of the 
surface $\mathrm{A}$. This can be done by using
the first four quadrics of Eq.(\ref{INTER}) 
to eliminate in a linear
way the variables $\overline{{\bf{c}}},
\overline{{\bf{d}}},
{\bf{f}}$ and $
\overline{{\bf{g}}}$. The remaining 
quartic $\mathrm{Q}_5$ gives rise to the surface
$\mathrm{A} \in [{\bf{a}}:{\bf{b}}:\overline{{\bf{b}}}:{\bf{g}}]$
whose defining polynomial can be written as,
\begin{equation}
\label{SUPA}
\mathrm{A}=\mathrm{F}_1^2-\frac{\mathrm{U}^2}{q}({\bf{a}} {\bf{g}}+{\bf{b}} \overline{{\bf{b}}}) \mathrm{F}_2,
\end{equation}
where the polynomials $\mathrm{F}_1$ and $\mathrm{F}_2$ are given by,
\begin{eqnarray}
\label{SUPAA}
&& \mathrm{F}_1 = ({\bf{a}}^2-{\bf{g}}^2)^2+{\bf{b}}^4+\overline{{\bf{b}}}^4-4{\bf{a}}{\bf{b}}\overline{{\bf{b}}}{\bf{g}}
-(q+\frac{1}{q})({\bf{a}}^2+{\bf{g}}^2)({\bf{b}}^2+\overline{{\bf{b}}}^2)
-(q^2+\frac{1}{q^2})(2{\bf{a}}{\bf{g}}+{\bf{b}}\overline{{\bf{b}}}){\bf{b}}\overline{{\bf{b}}}, \nonumber \\
&& \mathrm{F}_2 = (q+\frac{1}{q})({\bf{a}}^2+{\bf{g}}^2)({\bf{b}}^2+\overline{{\bf{b}}}^2){\bf{b}}\overline{{\bf{b}}}
+\left[{\bf{b}}^4+\overline{{\bf{b}}}^4+(4+q^2+\frac{1}{q^2}){\bf{b}}^2\overline{{\bf{b}}}^2\right]{\bf{a}}{\bf{g}}
-{\bf{b}}\overline{{\bf{b}}}({\bf{a}}^2-{\bf{g}}^2)^2. \nonumber \\
\end{eqnarray}

We have verified that on the subspace (\ref{SUBM}) of couplings
the octic surface defined by Eqs.(\ref{SUPA},\ref{SUPAA}) decomposes 
into lower degree polynomials. More specifically, we find that
for $\varepsilon=1$ such reducibility is in terms of
the product of two quartics surfaces while 
for $\varepsilon=-1$ we have a product of
four quadrics surfaces. In both cases all these 
surfaces components 
are birational to $\mathbb{CP}^2$ and thus 
rational varieties.

Interesting enough, we also see that there 
is another simple degeneration 
possibility once we set $\mathrm{U}=0$. In this case 
the surface $\mathrm{A}$ becomes a square of the 
polynomial $\mathrm{F}_1$ for arbitrary values of $q$.
It turns out that the quartic surface defined by 
$\mathrm{F}_1$ contains only simple 
singularities whose minimal resolution is known to be 
a $\mathrm{K3}$ surface \cite{BU}.

\section{The symmetric gauge geometry }

It has been noted in \cite{BEKO} that for 
a particular choice 
of the gauge parameter
$\gamma$ many of the off-diagonal amplitudes 
of the $\mathrm{R}$-matrix 
becomes symmetric under transposition. This occurs when,
\begin{equation}
\label{TRAVA}
\gamma^2 \sim \frac{\sqrt{q(\xi+x^{+})}(x^{+}-x^{-})}{\sqrt{(1+\xi^2)(\xi+x^{-})}}. 
\end{equation}

By substituting the relations (\ref{CHAN}) we conclude that Eq.(\ref{TRAVA}) is
equivalent to the condition that the ratio $z/w$ is constant. 
Clearly, this plane does not intersect the ruled surface $\mathrm{S}$ on its
$\mathbb{CP}^1$ bundle and consequently the geometric properties of the
Boltzmann weights may not be described by an elliptic curve. Without loss
of generality we can set $w=z$ for the symmetric gauge and the weights
are now sited in the following curve,
\begin{equation}
\label{CUR}
\overline{\mathrm{C}}=(x^2-\frac{y^2}{q})(x^2-qy^2)^2-\mathrm{U}xyz^2(x^2-qy^2)-z^4(x^2-q^3y^2).
\end{equation}

The above curve has three singular points one of them is an ordinary singularity
while the others behave as tacnodes. The latter singularities behave like double
point with only two tangent but having two branches, 
see for example \cite{WAL}. The geometric genus 
$g(\overline{\mathrm{C}})$ is computed considering such 
singularities deficiencies and the
result is,
\begin{equation}
g(\overline{\mathrm{C}})= \frac{5 \times 4}{2} -1-2\times2=5.
\end{equation}

In this situation the relationship among the curve $\overline{\mathrm{C}}$ 
with the original torus $\mathrm{E}_1$ is more severe than
an isogeny. Indeed, the relation among such curves coordinates on
the affine plane $z=1$ becomes,  
\begin{equation}
\label{MAPC}
x_{+}=-\xi-\frac{\sqrt{1+\xi^2}}{\sqrt{q}}\left(\frac{y}{x}\right)(x^2-qy^2),~~
x_{-}=-\xi-\frac{\sqrt{1+\xi^2}}{q^{3/2}}\left(\frac{x}{y}\right)(x^2-qy^2),
\end{equation}
given rise to a ramified mapping among curves explaining the 
drastic change on the genus. For some potential geometric degeneracies 
associated to the symmetric gauge see Appendix C.

Let us now turn our attention to the geometry properties of the
$\mathrm{R}$-matrix in the symmetric gauge.
The expressions for the respective 
matrix elements are obtained setting
${\bf{z}}_j=1$ in the previous relations (\ref{RMAG}). We observe
that we have two less independent matrix 
elements because of the identities
$\overline{{\bf{c}}}={\bf{c}}$ 
and $\overline{{\bf{d}}}={\bf{d}}$. Considering this fact it 
follows from Eqs.(\ref{INTER})
that the underlying variety $\mathrm{Z}$ 
is now defined by,
\begin{equation}
\label{YY1}
\mathrm{Z}=\{ 
({\bf{a}}:{\bf{b}}:\overline{{\bf{b}}}:{\bf{c}}:{\bf{d}}:
{\bf{f}}:{\bf{g}}:\overline{{\bf{g}}}) 
\in \mathbb{CP}^7~|~\overline{\mathrm{Q}}_1=\overline{\mathrm{Q}}_2=\overline{\mathrm{Q}}_3=\overline{\mathrm{Q}}_4=\overline{\mathrm{Q}}_5=0\},
\end{equation}
where the polynomials $\overline{\mathrm{Q}}_j$ are only quadrics of the form,
\begin{eqnarray}
\label{QUA1}
&&\overline{\mathrm{Q}}_1={\bf{b}}\overline{\bf{b}}+{\bf{a}}{\bf{g}}-{\bf{c}}^2,~~
\overline{\mathrm{Q}}_2=\overline{\bf{b}}{\bf{b}}+{\bf{f}}\overline{\bf{g}}+q{\bf{d}}^2, \nonumber \\
&&\overline{\mathrm{Q}}_3={\bf{g}}{\bf{f}}+{\bf{a}}\overline{\bf{g}}+(q+1/q){\bf{b}}\overline{\bf{b}},~~
\overline{\mathrm{Q}}_4={\bf{a}}{\bf{f}}+{\bf{g}}\overline{\bf{g}}-({\bf{b}}^2+\overline{\bf{b}}^2), \nonumber \\
&&\overline{\mathrm{Q}}_5=(\overline{{\bf{g}}}-q{\bf{g}})({\bf{g}}-q\overline{\bf{g}})
-({\bf{f}}-q{\bf{a}})({\bf{a}}-q{\bf{f}})
-q\mathrm{U}{\bf{c}}{\bf{d}}.
\end{eqnarray}

The algebraic set $\mathrm{Z}$ is a complete intersection and 
consequently we are dealing with a complex surface. One way to unveil its
geometrical invariants is through a mapping to another variety whose
geometric data is known. Here we are fortunate of being able 
to establish
a simple map to the Abelian surface of previous section. 
Considering the $\mathbb{CP}^3$ embeddings
for $\mathrm{A}$ and $\mathrm{Z}$
given by Eqs.(\ref{SUPA},\ref{SUPAA},\ref{SUP16},\ref{SUP16A})
we can set the mapping,
\begin{equation}
\renewcommand{\arraystretch}{1.5}
\begin{array}{ccc}
\mathrm{Z} \subset \mathbb{C}  
$[${\bf{a}}$,${\bf{b}}$,$\overline{\bf{b}}$,${\bf{c}}$]$
&~~~ \overset{\psi}{-~-~-\rightarrow}~~~ 
& \mathrm{A} \subset\mathbb{C} 
$[${\bf{a}}$,${\bf{b}}$,$\overline{\bf{b}}$,${\bf{g}}$]$
\\
$[${\bf{a}}$:${\bf{b}}$:$\overline{\bf{b}}$:${\bf{c}}$]$
& \longmapsto &~
$[${\bf{a}}^2$:${\bf{a}}{\bf{b}}$:${\bf{a}}\overline{\bf{b}}$:${\bf{c}}^2-{\bf{b}}\overline{{\bf{b}}}$]$.
\end{array}
\end{equation}

The map $\psi$ has degree two being regular in the open set ${\bf{a}}=1$ 
containing four
ramification lines at ${\bf{c}}=0$. From Hironaka desingularization 
theorem \cite{HIRO} it follows that by a succession of monoidal 
transformations we can eliminate the 
indeterminacy locus of $\psi$ resulting in
a morphism,
\begin{equation}
\renewcommand{\arraystretch}{1.5}
\begin{array}{ccc}
\widetilde{\mathrm{Z}} \subset \mathbb{CP}^3  
&~~~ \overset{\widetilde{\psi}} {\longrightarrow}~~~ 
& \widetilde{\mathrm{A}} \subset\mathbb{CP}^3, 
\end{array}
\end{equation}
connecting the birational models $\widetilde{Z}$ and $\widetilde{A}$ 
of the surfaces $\mathrm{Z}$ and $\mathrm{A}$, respectively.

Now the map $\widetilde{\psi}$ defines a double covering 
branched along 
the union of disjoint smooth curves with an effective locus say
$\mathrm{B} \in \widetilde{\mathrm{A}}$. Since
the work by Persson \cite{PER} it is known that a smooth double cover
of the surface $\widetilde{\mathrm{A}}$ is uniquely determined by 
a line bundle $\mathcal{L}$ on
$\widetilde{\mathrm{A}}$ such that $\mathrm{B} \in |2 \mathcal{L}|$.
For recent overview on the properties of the invariants of double 
coverings of surfaces
see for instance \cite{HUL}. It turns out that from this construction
we can uncover
the geometric data of $\widetilde{\mathrm{Z}}$ as follows,

$\bullet $ The Euler-Poincar\'e characteristic $\chi(\mathrm{S})$ of surface $\mathrm{S}$
\EQ
\label{EUL}
\chi(\widetilde{\mathrm{Z}})=2 \chi(\widetilde{\mathrm{A}})+\frac{1}{2} \left( \mathcal{L},\mathcal{L}+\mathrm{K}_{\widetilde{\mathrm{A}}} \right)
=\frac{1}{2}\left( \mathcal{L},\mathcal{L} \right),
\EN
where
$\mathrm{K}_{\mathrm{S}}$ is the canonical bundle of $\mathrm{S}$ and 
$\left ( \mathcal{L}, \mathrm{D} \right )$ denotes
the intersection number of the line bundle and a divisor $\mathrm{D} \in \widetilde{\mathrm{A}}$.

$\bullet $ The self-intersection number $\mathrm{K}_{\mathrm{S}}^2$ on the surface $\mathrm{S}$
\EQ
\label{KKINT}
\mathrm{K}_{\widetilde{\mathrm{Z}}}^2=
2\mathrm{K}_{\widetilde{\mathrm{A}}}^2+
2 \left (\mathcal{L} 
,\mathcal{L}+\mathrm{K}_{\widetilde{\mathrm{A}}} \right)
+2 \left (\mathcal{L} 
,\mathrm{K}_{\widetilde{\mathrm{A}}} \right)
=2\left( \mathcal{L},\mathcal{L} \right).
\EN

$\bullet $ The geometric genus $p_g(\mathrm{S})$ of the surface $\mathrm{S}$
\EQ
\label{PGG}
p_g(\widetilde{\mathrm{Z}})=p_g(\widetilde{\mathrm{A}})+ 
dim_{\mathbb{C}}H^0(\widetilde{\mathrm{A}},\mathrm{K}_{\widetilde{\mathrm{A}}}\otimes\mathcal{L})
=1+dim_{\mathbb{C}}H^0(\widetilde{\mathrm{A}},\mathcal{L})
\EN
where $H^{i}(\widetilde{\mathrm{A}},\mathcal{L})$ denotes the $i$-th cohomology group of the sheaf
associated to the line bundle $\mathcal{L}$.

$\bullet $ The irregularity $q(\mathrm{S})$ of the surface $\mathrm{S}$
\EQ
\label{IRRE}
q(\widetilde{\mathrm{Z}})=1+p_q(\widetilde{\mathrm{Z}})-\chi(\widetilde{\mathrm{Z}})=
2+dim_{\mathbb{C}}H^0(\widetilde{\mathrm{A}},\mathcal{L})
-\frac{1}{2}\left( \mathcal{L},\mathcal{L} \right)
\EN

The last equality in the above formulas considered 
the properties of an Abelian
surface, that is 
$\chi(\widetilde{\mathrm{A}})=0$ and $p_q(\widetilde{\mathrm{A}})=1$ 
as well as the fact
$\mathrm{K}_{\widetilde{\mathrm{A}}}$ is trivial.
At this point we 
see from Eqs.(\ref{EUL},\ref{KKINT}) that the 
corresponding invariants satisfy the
relation,
\EQ
\mathrm{K}_{\widetilde{\mathrm{Z}}}^2=4 
\chi(\widetilde{\mathrm{Z}})
\EN
which means that the surface 
$\widetilde{\mathrm{Z}}$ sits on 
the so-called Severi line \cite{SE}.

From now on we assume that $\mathcal{L}$ is an ample 
line bundle and under this
mild hypothesis we can relate the dimension of
$H^0(\widetilde{\mathrm{A}},\mathcal{L})$ with the
self-intersection number
of the line bundle. Indeed, from the Riemann-Roch theorem \cite{BU}
for the Euler-Poincar\'e
characteristic of $\mathcal{L}$ we have,
\EQ
\label{RRC}
dim_{\mathbb{C}}H^0(\widetilde{\mathrm{A}},\mathcal{L})
-dim_{\mathbb{C}}H^1(\widetilde{\mathrm{A}},\mathcal{L})
+dim_{\mathbb{C}}H^2(\widetilde{\mathrm{A}},\mathcal{L})=
\frac{1}{2} \left( \mathcal{L},\mathcal{L} \right)
\EN
and considering that the 
ampleness assumption for the line bundle implies
$dim_{\mathbb{C}}H^1(\widetilde{\mathrm{A}},\mathcal{L})
=dim_{\mathbb{C}}H^2(\widetilde{\mathrm{A}},\mathcal{L})=0$ 
we obtain from Eq.(\ref{RRC}) 
the simple relation,
\EQ
dim_{\mathbb{C}}H^0(\widetilde{\mathrm{A}},\mathcal{L})
=\frac{1}{2} \left( \mathcal{L},\mathcal{L} \right)
\EN

Note that the above relation together with Eq.(\ref{IRRE}) 
already fix the 
irregularity of $\widetilde{\mathrm{Z}}$, namely
\EQ
q(\widetilde{\mathrm{Z}})=2
\EN

The complete understanding of the geometric properties 
of $\widetilde{Z}$ still needs the knowledge
of the line bundle intersection number. We can retrieve this number
with the help of the adjunction formula for a generic curve $\widetilde{\mathrm{C}}$
on $\widetilde{\mathrm{A}}$, namely
\EQ
\label{ADJ}
g(\widetilde{\mathrm{C}})=1+
\frac{1}{2} \left( \widetilde{\mathrm{C}},\widetilde{\mathrm{C}} \right)
\EN

We stress here that the self-intersection 
number for 
$\widetilde{\mathrm{C}}$ is defined to be the 
intersecting number of the
corresponding line bundle. 
Hence for any curve
$\widetilde{\mathrm{C}}$ in the linear system
of the ample line bundle $\mathcal{L}$ 
we can reverse Eq.(\ref{ADJ}) and state that,
\EQ
\left( \mathcal{L},\mathcal{L} \right)=
2[g(\widetilde{\mathrm{C}})-1]
\EN

We now are left to compute the genus of 
an ample divisor on the
original Abelian surface such 
as the intersection of $\mathrm{A}$ 
with a hyperplane in some 
projective embedding. 
For this purpose it is suffice to 
take the curve
obtained from 
Eq.(\ref{SUPA},\ref{SUPAA}) when
we set for instance ${\bf{g}}=0$,
\begin{eqnarray}
\mathrm{C} &=& \left[ {\bf{a}}^4+{\bf{b}}^4+\overline{{\bf{b}}}^4
-(q+\frac{1}{q}){\bf{a}}^2({\bf{b}}^2+\overline{{\bf{b}}}^2)
-(q^2+\frac{1}{q^2})({\bf{b}}\overline{{\bf{b}}})^2 \right]^2 \nonumber \\
  &-& \frac{\mathrm{U}^2}{q}({\bf{a}}{\bf{b}}\overline{{\bf{b}}})^2\left[
(q+\frac{1}{q})({\bf{b}}^2+\overline{{\bf{b}}}^2)- {\bf{a}}^2 \right]
\end{eqnarray}

It turns out that such degree eight plane curve 
has twelve ordinary 
singularities and all of them have the multiplicity 
index of double points. The genus
of its desingularization $\widetilde{\mathrm{C}}$
is therefore easily computed as,
\EQ
g(\widetilde{\mathrm{C}})=\frac{7 \times 6}{2}-12=9
\EN
and consequently we obtain 
$\left( \mathcal{L},\mathcal{L} \right)=16$.

Now, collecting together all the above information we 
conclude that
the surface we have started with is of general type whose main
geometric data is,
\EQ
\label{INVT}
q(\widetilde{\mathrm{Z}})=2~~\mathrm{and}~~p_q(\widetilde{\mathrm{Z}})=9
\EN

In addition to that,
considering the values 
$\chi(\widetilde{\mathrm{Z}})=8$ and $\mathrm{K}_{\widetilde{\mathrm{Z}}}^2=32$ we
can predict the behaviour of the 
corresponding higher $n$-plurigenera, namely
\EQ
P_n(\widetilde{\mathrm{Z}})= \chi(\widetilde{\mathrm{Z}}) +\frac{n(n-1)}{2} \mathrm{K}_{\widetilde{\mathrm{Z}}}^2=
8\left[1+2n(n-1)\right]~~\mathrm{for}~~n>2
\EN

Finally, we remark that we have checked some of 
these numerical values
within the formal surface desingularization routine 
implemented in the
computer algebra system Magma \cite{MAG}. Not only we have been
able to confirm the values for irregularity and the geometric
genus but also the first 
two $P_2(\widetilde{\mathrm{Z}})$ and $P_3(\widetilde{\mathrm{Z}})$ plurigenera numbers.

\section{Conclusions}

In this work, we have derived a formulation 
for the $\mathrm{R}$-matrix 
based on a $q$-deformation of the centrally extended $sl(2|2)$ 
superalgebra towards the view of algebraic geometry. This made it
possible to uncover the geometric properties of the elementary
weights as well as of the corresponding $\mathrm{R}$-matrix.

Our analysis made clear that if $\mathrm{X}$ denotes the variety 
associated to the elementary weights the geometry underlying the $\mathrm{R}$-matrix
is not necessarily described by the product 
$\mathrm{X} \times \mathrm{X}$. We have argued that this change is rather drastic
for the symmetric choice of the gauge parameter. In fact,
in this case $\mathrm{X}$ is the genus five curve 
$\overline{\mathrm{C}}$ (\ref{CUR}) and the geometric data of the product 
$\overline{\mathrm{C}} \times \overline{\mathrm{C}}$ can be retrieved
directly from the genus of the respective curve.
Though this also gives rise to a surface of
general type the geometrical invariants are,
\EQ
q(\overline{\mathrm{C}} \times \overline{\mathrm{C}})=5+5=10~~~\mathrm{and}~~~
p_g(\overline{\mathrm{C}} \times \overline{\mathrm{C}})=5\times5=25
\EN
which are quite distinct from the actual geometrical content
associated to the $\mathrm{R}$-matrix, see Eq.(\ref{INVT}).

We have observed that the geometric properties can 
change to other classes of universalities when the
couplings are restricted to the subspace (\ref{SUBM}).
It is reasonable to think that such geometry change will
reflect on an equivalent modification of
the physical properties associated to the 
respective vertex model and spin chain. It seems
worthwhile to investigate the way the 
geometric properties may be encoded for instance
on the nature of the excitations of the spin
chain. It seems also of interest to carry on
the algebraic Bethe ansatz for both the 
eigenvectors and eigenvalues 
of the transfer matrix. 
We expect that the derived identities (\ref{QUA1})  
among the $\mathrm{R}$-matrix entries
will be useful for such algebraic
formulation.

\section*{Acknowledgments}
I thank N. Beisert for fruitful discussions 
and for the hospitality
at the Institute for Theoretical Physics, Zurich, where 
this work has been started.
This work was supported in part by 
the Brazilian Research Agency 
CNPq(2013/30329), the Pauli Center for Theoretical Studies 
and the Swiss National
Science Foundation NCCR SwissMap.

\addcontentsline{toc}{section}{Appendix A}
\section*{\bf Appendix A: Uniformization of $\mathrm{E}_2$}
\setcounter{equation}{0}
\renewcommand{\theequation}{A.\arabic{equation}}

We note that the curve $\mathrm{E}_2$ can be 
rewritten as, 
\begin{equation}
\label{EE2}
-\frac{y_1^2}{4q}=(1-\lambda_1 y_2^2)(1-\lambda_2 y_2^2),
\end{equation}
where the parameters $\lambda_1$ and $\lambda_2$ satisfy the 
following relations,
\begin{equation}
\lambda_1+\lambda_2=\frac{4+4q^4-q\mathrm{U}^2}{4q},~~~\lambda_1 \lambda_2=q^2.
\end{equation}

Rescaling the variables $y_1 \rightarrow 2\IM\sqrt{q}y_1$ and 
$y_2 \rightarrow y_2/\sqrt{\lambda_1}$ we can bring Eq.(\ref{EE2}) in
the standard Jacobi form. A natural uniformization is therefore
in terms of the Jacobi's elliptic functions, namely
\begin{equation}
y_1= 2 \IM \sqrt{q}\mathrm{cn}(\mu,{\bf{k}}) \mathrm{dn}(\mu,{\bf{k}}),~~~y_2= \sqrt{\frac{{\bf{k}}}{q}}\mathrm{sn}(\mu,{\bf{k}}),
\end{equation}
where $\mu$ is a spectral parameter and ${\bf{k}}$ is the modulus 
of the elliptic functions. The latter
is given in terms of the couplings $q$ and $\mathrm{U}$ by,
\begin{equation}
{\bf{k}}= \frac{\Delta}{2} \pm \sqrt{\Delta^2/4-1},~~~\Delta=q^2 +\frac{1}{q^2}-\frac{\mathrm{U}^2}{4q}.
\end{equation}

\addcontentsline{toc}{section}{Appendix B}
\section*{\bf Appendix B: Surface Analysis}
\setcounter{equation}{0}
\renewcommand{\theequation}{B.\arabic{equation}}
After extracting the product  
${\bf{b}} \overline{{\bf{b}}}$ from quadric $\mathrm{Q}_3$ and substituting it in Eq.(\ref{Q5}) we obtain,
\begin{eqnarray}
\widetilde{\mathrm{Q}}_5&=&\left[(\overline{{\bf{g}}}-q{\bf{g}})({\bf{g}}-q\overline{\bf{g}})
-({\bf{f}}-q{\bf{a}})({\bf{a}}-q{\bf{f}})\right]^2 \nonumber \\
&+& \frac{q^3\mathrm{U}^2}{(1+q^2)^2} \left[{\bf{f}}(q\overline{{\bf{g}}}-{\bf{g}})+\overline{{\bf{g}}}({\bf{f}}/q-{\bf{a}})\right]
\left[{\bf{a}}(q{\bf{g}}-\overline{{\bf{g}}})-{\bf{g}}({\bf{f}}-{\bf{a}}/q)\right].
\end{eqnarray}

The geometry of the above quartic surface can be understood 
by means of composition of birational transformations. We start by 
defining the following
auxiliary variables,
\begin{equation}
{\bf{h}}={\bf{a}}-{\bf{f}}/q,~~~
\overline{{\bf{h}}}={\bf{a}}-q{\bf{f}},~~~
{\bf{p}}={\bf{g}}-\overline{{\bf{g}}}/q,~~~
\overline{{\bf{p}}}={\bf{g}}-q\overline{{\bf{g}}}.
\end{equation}

We then observe that $\widetilde{\mathrm{Q}}_5$ becomes 
quadratic in the variable 
$\overline{{\bf{p}}}$ and the linear term can be eliminated 
by quadrature. More precisely, we are able
to perform the following
transformation,
\begin{equation}
\frac{\overline{{\bf{p}}}}{{\bf{h}}}=\frac{\IM\sqrt{q}(q^4-1)\mathrm{U}x_0x_1-\left[q(1+q^4)\mathrm{U}^2-2(1-q^4)^2\right]{\bf{p}}\overline{{\bf{h}}}}{2\left[(q^4-1)^2{\bf{p}}^2-q\mathrm{U}^2\overline{{\bf{h}}}\right]},
\end{equation}
replacing the variables 
${\overline{{\bf{p}}}}$ and ${{\bf{h}}}$ by the new homogeneous coordinates $x_0$ and $x_1$.

As a result we find that the expression of the surface $\widetilde{\mathrm{Q}}_5$ in terms of these new
variables is,
\begin{equation}
\widetilde{\mathrm{Q}}_5=x_0^2 x_1^2+4q^4{\bf{p}}^4-(4-q\mathrm{U}^2+4q^4){\bf{p}}^2\overline{{\bf{h}}}
+4\overline{{\bf{h}}}^4,
\end{equation}
which is exactly the ruled surface (\ref{SUP}) upon re-scaling of the coordinates 
${\bf{p}}=\frac{x_2}{q^{3/4}}$ and
$\overline{{\bf{h}}}=q^{3/4}x_3$.

The last step in the analysis concerns with the study of the 
curve originated
from the polynomial $\mathrm{Q}_4$. Taking into account 
the above information we find
that it has the following structure,
\begin{eqnarray}
\mathrm{Q}_4&=&{\bf{b}}^4+\mathrm{P}_1(y_1,y_2){\bf{b}}^2x_2^2+ 
\mathrm{P}_2(y_1,y_2){\bf{b}}^2x_2+ 
\mathrm{P}_3(y_1,y_2){\bf{b}}+ 
\mathrm{P}_4(y_1,y_2)x_2^4+ 
\mathrm{P}_5(y_1,y_2)x_2^3\nonumber \\ 
&+&\mathrm{P}_6(y_1,y_2)x_2^2+ 
\mathrm{P}_7(y_1,y_2)x_2+ 
\mathrm{P}_8(y_1,y_2), 
\end{eqnarray}
where the coefficients $\mathrm{P}_j(y_1,y_2)$ belong 
to the field of fractions of $\mathrm{E}_2$.

We find that this curve has two ordinary double points 
as singularities and therefore its
normalization $\mathrm{E}_3$ is an elliptic curve. Remarkably, the respective
$\mathrm{J}$-invariant does depend on the curve $\mathrm{E}_2$ and its explicit value is,
\begin{equation}
\mathrm{J}(\mathrm{E}_3)=
\frac{\left(16-8q\mathrm{U}^2+q^2\mathrm{U}^4+960q^2-240q^3\mathrm{U}^2+2144q^4-8q^5\mathrm{U}^2+960q^6+16q^8\right)^3}{q^2(4-q\mathrm{U}^2+4q^4+8q^2)(4-q\mathrm{U}^2+4q^4-8q^2)^4}.
\end{equation}

\addcontentsline{toc}{section}{Appendix C}
\section*{\bf Appendix C: Symmetric Gauge Degenerations }
\setcounter{equation}{0}
\renewcommand{\theequation}{C.\arabic{equation}}

In this case the genus five curve (\ref{CUR}) 
decomposes into two 
irreducible and isomorphic elliptic curves
provided the constrain (\ref{SUBM}) 
is satisfied. The expression 
of one of the components is,
\begin{equation}
\overline{\mathrm{C}}=x^3+x^2y/\sqrt{q}-qxy^2-\varepsilon 
\sqrt{q}y^3+\varepsilon xz^2-q^{3/2}yz^2.
\end{equation}

Interesting enough, we note that the  
the two possible cases for $\varepsilon$ are distinguished by the
$\mathrm{J}$-invariant,
\begin{displaymath}
\mathrm{J}(\overline{\mathrm{C}})= \left \{ \begin{array}{cc}
\frac{64(q^2+3)^3(3q^2+1)^3}{(q^2-1)^4(q^2+1)^2} & \mathrm{for}~~\varepsilon=1 \\ 
1728 & \mathrm{for}~~\varepsilon=-1 . 
\end{array}
\right.
\end{displaymath}

As in the main text we can embed 
the variety $\mathrm{Z}$ in a $\mathbb{CP}^3$ projective space.
For $\mathrm{U} \neq 0$ we can use the quadrics (\ref{QUA1}) to
eliminate the elements 
${\bf{d}},{\bf{f}},{\bf{g}},\overline{{\bf{g}}}$ and such
embedding is given by,
\begin{equation}
\label{SUP16}
\mathrm{Z}=\mathrm{F}_3^2-\frac{\mathrm{U}^2}{q}{\bf{a}}^4{\bf{c}}^2 \mathrm{F}_4,
\end{equation}
where the polynomials $\mathrm{F}_3$ and $\mathrm{F}_4$ are,
\begin{eqnarray}
\label{SUP16A}
\mathrm{F}_3 & =& ({\bf{a}}^2+{\bf{c}}^2-{\bf{b}}\overline{{\bf{b}}})^2({\bf{a}}^2-{\bf{c}}^2+{\bf{b}}\overline{{\bf{b}}})^2+{\bf{a}}^4\left[{\bf{b}}^4+\overline{{\bf{b}}}^4-4{\bf{b}}\overline{{\bf{b}}}({\bf{c}}^2-{\bf{b}}\overline{{\bf{b}}})\right] \nonumber \\
&-&(q+\frac{1}{q}){\bf{a}}^2({\bf{b}}^2+\overline{{\bf{b}}}^2)\left[{\bf{a}}^4+({\bf{c}}^2-{\bf{b}}\overline{{\bf{b}}})^2\right]
+(q^2+1/q^2){\bf{a}}^4{\bf{b}}\overline{{\bf{b}}}({\bf{b}}\overline{{\bf{b}}}-2{\bf{c}}^2), \nonumber \\
\mathrm{F}_4 &=& (q+\frac{1}{q}){\bf{a}}^2{\bf{b}}\overline{{\bf{b}}}({\bf{b}}^2+\overline{{\bf{b}}}^2)\left[{\bf{a}}^4+({\bf{c}}^2-{\bf{b}}\overline{{\bf{b}}})^2\right]
+(4+q^2+\frac{1}{q^2}){\bf{a}}^4{\bf{b}}^2\overline{{\bf{b}}}^2\left[{\bf{c}}^2-{\bf{b}}\overline{{\bf{b}}}\right] \nonumber \\
&-&{\bf{b}}\overline{{\bf{b}}}({\bf{a}}^2+{\bf{c}}^2-{\bf{b}}\overline{{\bf{b}}})^2({\bf{a}}^2-{\bf{c}}^2+{\bf{b}}\overline{{\bf{b}}})^2+{\bf{a}}^4({\bf{b}}^4+\overline{{\bf{b}}}^4)({\bf{c}}^2-{\bf{b}}\overline{{\bf{b}}}).
\end{eqnarray}

For $\varepsilon=1$ the surface (\ref{SUP16},\ref{SUP16A}) 
factorizes into the product of two octic
surfaces which can be seen as a ramified double 
cover over $\mathrm{CP}^2$ and thus their
normalization are $\mathrm{K3}$ surfaces. On the other hand for $\varepsilon=-1$ we have 
a factorization on the complex field in terms of eight quadrics and consequently $\mathrm{Z}$ 
is described by rational surfaces.

\end{document}